\documentclass[12pt,preprint]{aastex}

\begin{document}

\title{THE FRACTAL DIMENSION OF PROJECTED CLOUDS}

\author{N\'estor S\'anchez,\altaffilmark{1,2}
        Emilio J. Alfaro,\altaffilmark{1} and 
        Enrique P\'erez\altaffilmark{1}}

\altaffiltext{1}{Instituto de Astrof\'{\i}sica de Andaluc\'{\i}a, CSIC,
                 Apdo. 3004, E-18080, Granada, Spain;
                 nestor@iaa.es, emilio@iaa.es, eperez@iaa.es.}
\altaffiltext{2}{Departamento de F\'{\i}sica, Universidad del Zulia,
                 Maracaibo, Venezuela.}

\begin{abstract}
The interstellar medium seems to have an underlying
fractal structure which can be characterized through
its fractal dimension. However, interstellar clouds
are observed as projected two-dimensional images, and
the projection of a tri-dimensional fractal distorts
its measured properties. Here we use simulated
fractal clouds to study the relationship between
the tri-dimensional fractal dimension ($D_f$) of
modeled clouds and the dimension resulting from
their projected images. We analyze different
fractal dimension estimators: the correlation and
mass dimensions of the clouds, and the perimeter-based
dimension of their boundaries ($D_{per}$). We find
the functional forms relating $D_f$ with the projected
fractal dimensions, as well as the dependence on the
image resolution, which allow to estimate the ``real"
$D_f$ value of a cloud from its projection. The
application of these results to Orion A indicates
in a self-consistent way that $2.5 \la D_f \la 2.7$
for this molecular cloud, a value higher than the result
$D_{per}+1 \simeq 2.3$ some times assumed in literature
for interstellar clouds.
\end{abstract}


\section{INTRODUCTION}

Maps of nearby cloud complexes have shown that the
gas and dust are organized into irregular hierarchical
structures which seem similar to each other over a wide
range of scales \citep{scal90,falg92}. This self-similarity
has been interpreted as a signature of an underlying fractal
geometry that may be related to the physical processes
supporting the generation of structures in the interstellar
medium. The origin of this fractal-like structure seems to be
turbulence \citep{falg90,krit04,pado04}, although self-gravity
could also play an important role \citep{deve96}. A fractal
structure for the interstellar medium is consistent with many
observed features, including the cloud size and cloud mass
distribution functions \citep{elme96}, the properties of the
intercloud medium \citep{elme97a}, and even the stellar initial
mass function \citep{lars92,elme97b,elme02}.

Therefore, it is important to measure the degree of fractality
and/or complexity of the ISM. There are several ways this can
be done, such as structure tree methods \citep{houl92}, delta
variance techniques \citep{stut98}, principal components analysis
\citep{ghaz99}, multifractal analysis \citep{chap01}, metric space
techniques \citep{khal04}, etc. A simple method consists of
calculating the fractal dimension characterizing structures in
the ISM, but it has to be mentioned that quantifying the fractal
dimension of an object is generally not sufficient to characterize
it, because objects with different morphological properties could
have the same fractal dimension \citep{mand83}. In spite of this
a number of studies have been done to estimate the fractal dimension
of interstellar clouds. Most of these studies use the so-called
perimeter-area method: if the iso-contours exhibit a power-law
perimeter-area relation with a noninteger exponent over certain
range of scales, this exponent may be interpreted in terms of a
fractal dimension ($D_{per}$) that characterizes the manner these
curves fill space \citep{mand83}. This method was used by
\citet{beec87} on a group of selected dark clouds to estimate
a fractal dimension of $1.4$, whereas \citet{baze88} calculated
an average fractal dimension of $1.26$ for cirrus regions discovered
by IRAS. \citet{dick90} studied five different molecular cloud
complexes (Chamaeleon, R CrA, $\rho$ Oph, Taurus and the Lynds
134/183/1778 group) finding $1.17 \la D_{per} \la 1.28$. Also for
the Taurus complex, \citet{scal90} found $D_{per}=1.4$ and
\citet{falg91} found $D_{per}=1.36$ over a very wide range of
sizes. \citet{hete93} reported $1.3 \la D_{per} \la 1.52$ for
the Chamaeleon complex with an average of $1.44$, a value
notoriously higher than the one of \citet{dick90}. \citet{voge94}
also estimate relatively high values for $D_{per}$ which generally
range between $1.35$ and $1.5$ for high-velocity clouds and IRAS
cirrus.

The observational evidence can be summarized by saying that the
boundaries of molecular clouds appear to be fractal curves with
dimension $D_{per} \sim 1.35$ (or some value in the range
$1.2 \la D_{per} \la 1.5$). But the clouds are necessarily
recorded as two-dimensional images projected onto a plane
perpendicular to the line-of-sight. The connection between
$D_{per}$ and the fractal dimension of the tri-dimensional
clouds is still an open question. The fractal nature of the
projected boundaries suggests that these clouds may have fractal
surfaces with dimensions given by $D_{per}+1$ \citep{mand83}. Then,
it is usually assumed that $D_{per}+1 \simeq 2.35$ should be the
fractal dimension of interstellar clouds \citep[e.g.][]{beec92}.
However, although it is possible that much of their internal
structure is directly reflected in surface features, there is
not necessarily any simple relation between the fractal dimension
of the surface of a cloud and that of its internal structure. On
the other hand, it has been proven that the intersection of a
fractal with dimension $D_f$ and a plane gives another fractal
with dimension given by $D_f-1$ \citep{mand83}, but a projection
is a totally different operation and the result $D_{pro}=D_f-1$
does not have to be expected for the projected fractal dimension.
In fact, if an object of fractal dimension $D_f$ embedded in a
tri-dimensional Euclidean space, is projected on a plane, it is
possible to show that the projection has dimension $D_{pro}=D_f$
or $D_{pro}=2$ depending on whether $D_f<2$ or $D_f >2$, respectively
\citep{falc90}. Thus, the projection of a cloud with $D_f \simeq 2.35$
would give rise to a compact shadow that may not be a fractal,
but we want to emphasize that the relationship between the
fractal dimension of this projection and that of its boundary
is not clearly established.

The purpose of this work is to investigate the relationship
between the fractal dimension of a tri-dimensional cloud and
the fractal dimension of its projection, both for the whole
projected image and for its boundary. To do this we use
artificial fractal clouds generated by using an algorithm
which mimics in some way the hierarchical fragmentation
process occurring in molecular cloud complexes. Our main
goal is to estimate the real fractal dimension of
interstellar clouds from observed images. In \S~\ref{theory},
we explain the manner in which we simulate the clouds and,
after a brief review on some definitions of fractal dimension,
we calculate the fractal (correlation and mass) dimensions of
the projected clouds. In \S~\ref{perimeter}, we address the
perimeter-area relation of the projected images and its
dependence on the image resolution. This analysis is applied
to the Orion A molecular cloud in \S~\ref{orionA} and, finally,
the main results are discussed in \S~\ref{conclusions}.

\section{FRACTAL DIMENSION OF SIMULATED CLOUDS}
\label{theory}

\begin{figure*}
\centering
\includegraphics[width=17cm]{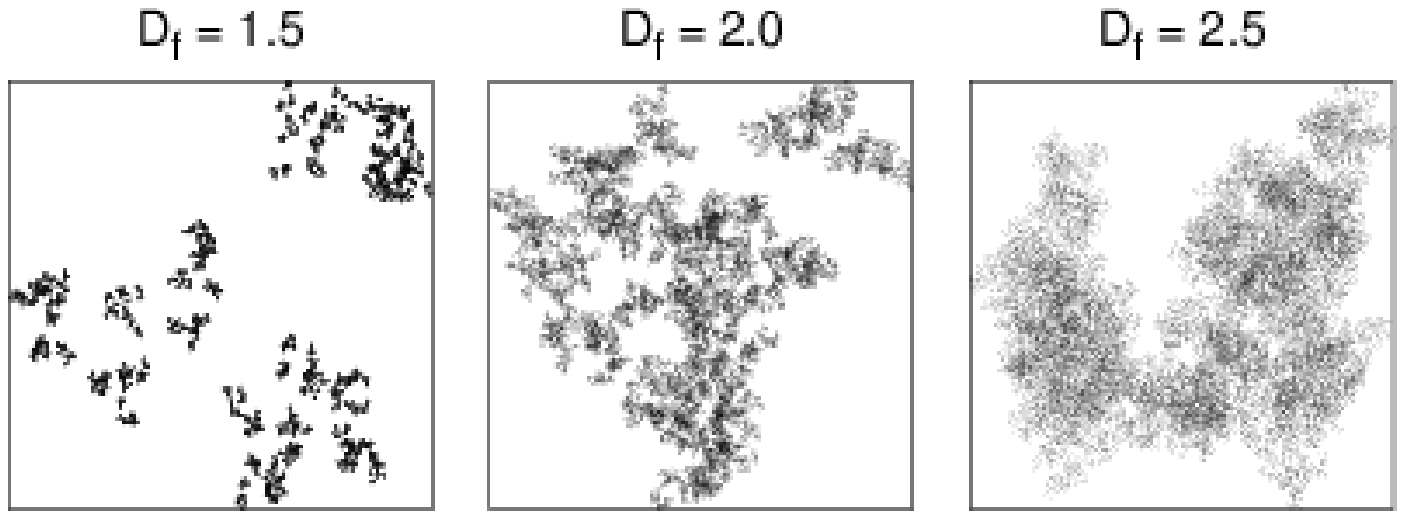}
\caption{Examples of clouds simulated with $N=3$ fragments
         per step in $H=9$ levels of hierarchy for three
         different values of fractal dimension: $1.5$, $2.0$
         and $2.5$.}
\label{clouds}
\end{figure*}

We first have to generate tri-dimensional fractal clouds.
For convenience we use simple recursive construction rules
based on the \citet{sone78} method. Within a sphere of radius
$R$ we place randomly $N$ spheres of radius $R/L$ with $L>1$,
in each of these spheres we place again $N$ smaller spheres with
radius $R/L^2$, and so on up to level $H$ of hierarchy. The set of
$N^H$ points (actually little spherical particles of radius
$R/L^H$) in the last level forms an object with fractal
dimension given by $D_f=\log{N}/\log{L}$. Most simulations
were made using $N=3$ fragments in a total of $H=9$ levels
of hierarchy ($\sim 2\times 10^4$ particles in the last level)
with the fractal dimension in the range $1 < D_f < 3$. However,
when the fractal dimension is too high ($D_f \ga 2.6$) the
filling factor tends to $1$ and it becomes difficult to place
randomly the spheres through all the volume without they being
superposed\footnote{We avoid superposition because the tests
made allowing superposition showed that the properties are 
modified in such a way that it appears a multifractal behavior
and a unique fractal dimension can not be defined in a wide
scale range}. In these cases we use $N=8$ fragments in $H=5$
levels ($\sim 3\times 10^4$ particles) and each fragment is placed
randomly {\it within} the available volume in the direction of
each of the eight octants. This allows us to reach high fractal
dimension values (and filling factors) easily. In any case,
several tests showed that our results do not depend
(at least within the error bars) on $N$ and/or $H$ (as long as
we have a sufficiently high number of particles in the last
level). In order to illustrate the appearance of the point
distributions, Figure~\ref{clouds}
shows examples of projections of tri-dimensional fractals
generated by using $N=3$, $H=9$ and three different values
of fractal dimension ($D_f=1.5$, $2.0$ and $2.5$).

\subsection{Fractal Dimension Estimation}

In the following, we will focus our attention on the empirical 
determination of the fractal dimension both for the simulated
clouds and for their two-dimensional projections. In general,
a fractal can be defined as a set of points whose Hausdorff
dimension $D_H$ is strictly larger than its topological
dimension $D_T$ \citep{mand83}. To estimate the Hausdorff
dimension authors normally use working definitions which fit
their method and needs, and thus there is not a unique
definition of fractal dimension. In general, a fractal
quantity is a number which is connected to some length $l$
in a manner like $A \sim l^{D_A}$, where $D_A$ would be the
(constant) dimension associated to the quantity $A$. The most
straightforward way to estimate the Hausdorff dimension of a
fractal set is through the box-counting method, based on the
fact that the number of boxes, $N(r)$, having side $r$ needed
to cover a fractal object varies as $r^{-D_B}$, being $D_B$ an
estimation of $D_H$ \citep{mand83} which is called the
``box-counting dimension" (or ``capacity"). Then, if we
cover the object with a grid and count the number $N(r)$
of occupied cells of size $r$, we can compute
\begin{equation}
D_B = - \lim_{r\rightarrow 0} \frac{\log{N(r)}}{\log{r}}\ .
\end{equation}
In practice $D_B$ corresponds with the slope of the best fit
in a $\log {N}-\log{r}$ plot, but obviously the range where this
scale law is valid in real (finite) fractals is limited to the
range between the smallest measurable region (the resolution)
and the object size.

The major problem with the box-counting method is that it leads
to imprecise results because of its sensitivity to, for example,
the placement and orientation of the grid and the range and
sequence of values of box side length \citep{bucz98,gonz00}. A
more useful method when dealing with a point distribution in
space has been introduced by \citet{gras83}: given a set of
$N_p$ points with positions ${\bf x}$, one first computes the
correlation integral $C(r)$ defined as
\begin{equation}
C(r)=\frac{2}{N_p(N_p-1)} \sum_{1 \le i < j \le N_p}
H \left( r- | {\bf x}_i - {\bf x}_j | \right)\ ,
\end{equation}
where $H$ is the Heaviside step function. The summation
counts the number of pairs for which the distance
$| {\bf x}_i - {\bf x}_j |$ is less than $r$. For a
fractal set the correlation integral scales at small
$r$ like $r^{D_C}$, being $D_C$ the ``correlation
dimension" of the set, so that
\begin{equation}
D_C = \lim_{r\rightarrow 0} \frac{\log{C(r)}}{\log{r}}\ .
\end{equation}
The dimension $D_C$ can be identified with the slope of the
log-log plot of $C(r)$ versus $r$, but usually this power-law
behavior is valid only in a finite range. Different (larger)
slopes can be observed at very small scales which are associated
with random (experimental) errors, noise and with the lack of
statistics of finite samples at very small scales; and there are
deviations due to nonlinear effects for too large $r$ values
\citep{thei87}.

If one now regards the mass (number of particles) inside a sphere
of radius $r$ we can define the ``mass dimension" ($D_M$) if the
mass obeys the self-similar scaling relation \citep{mand83}:
\begin{equation}
M(r) \sim r^{D_M}\ .
\end{equation}
A simple algorithm to estimate $D_M$ consists in generating a
sequence of spheres with different radii, count the number of
points inside each one, and calculate the slope in a $\log{M}$
versus $\log{r}$ plot. An obvious problem is where to place the
spheres, because large fluctuations in $M$ (lower values) can
appear if the spheres lie by chance in or near the voids that
exist in the distribution. To avoid this under-sampling problem
we modified this algorithm in the following way. We choose
randomly a position within the fractal cloud where we place
spheres having progressively greater radii and we calculate the
density (number of points per unit volume) in each case. Usually
the density shows an absolute maximum corresponding to the
sphere which gives a better sampling in that point, then we
take only this sphere and reject all the other ones. This
procedure is repeated many times at random positions through
the whole fractal volume. Finally, if several spheres have
the same radius we only consider the densest sphere (i.e.,
the ``best" sampling) to calculate the mass corresponding to
that radius value. With this strategy we improve the sampling
taking the spheres containing more information throughout all
the structure.

\begin{figure}
\plotone{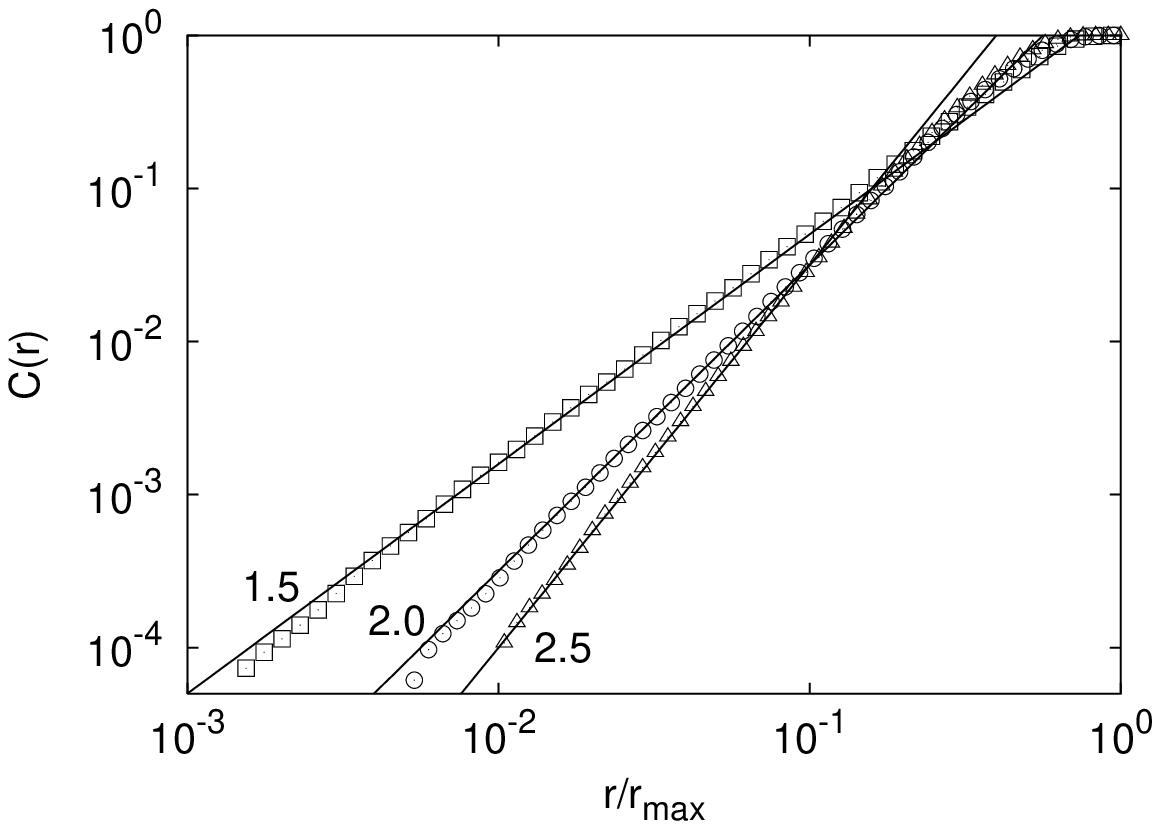}
\caption{The correlation integral $C(r)$
         for the three fractals shown in
         Figure~\ref{clouds}: $D_f=1.5$ (open squares),
         $D_f=2.0$ (open circles) and $D_f=2.5$ (open
         triangles). The lines show the expected slope
         for each fractal dimension value.}
\label{correlation3D}
\end{figure}

\begin{figure}
\centering
\plotone{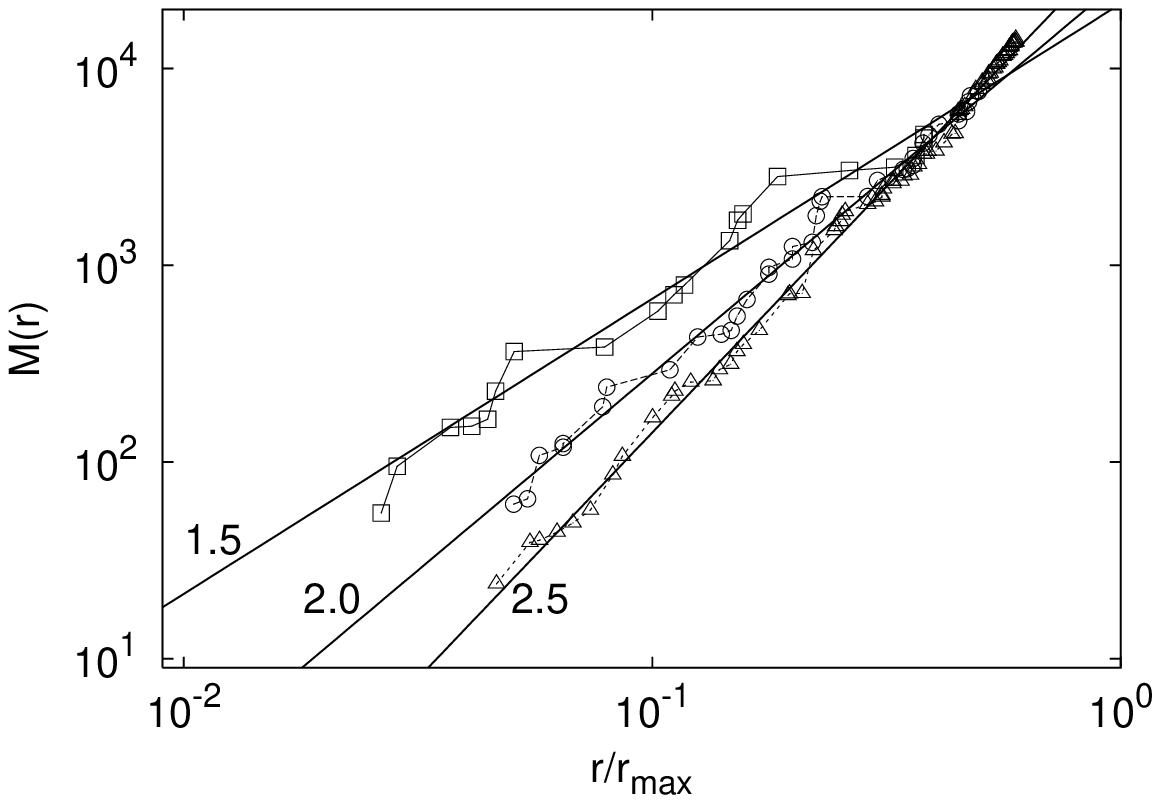}
\caption{The mass $M(r)$ of the spheres of radius $r$
         for the three fractals shown in Figure~\ref{clouds}:
         $D_f=1.5$ (open squares), $D_f=2.0$ (open circles) and
         $D_f=2.5$ (open triangles). The lines show the expected
         slope for each fractal dimension value.}
\label{mass3D}
\end{figure}

\begin{figure}
\centering
\plotone{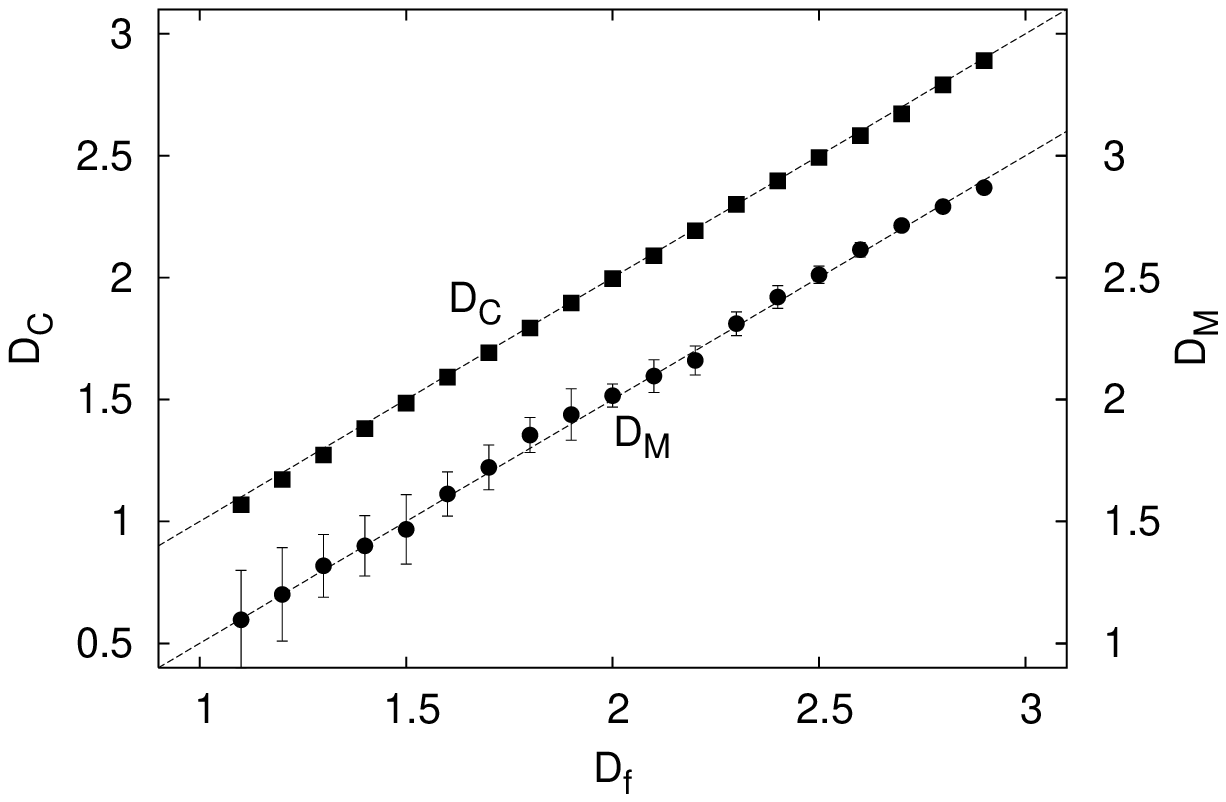}
\caption{The calculated correlation dimension $D_C$ (squares on
         the upper line and left side axis) and mass dimension
         $D_M$ (circles on the lower line and right side axis)
         as a function of the fractal dimension $D_f$ used to
         generate the cloud. Each point is the average of ten
         different realizations and the bars show the standard
         deviation. Dashed lines have slope unity.}
\label{both3D}
\end{figure}

We have calculated the correlation and mass dimensions as a
function of $D_f$ for the simulated tri-dimensional fractal
objects. Figure~\ref{correlation3D}
shows the correlation integral as a function of distance
(normalized to the maximum distance between two particles
in the fractal $r_{max}$) for the same three fractal clouds
shown in Figure~\ref{clouds}. The solid lines do not show the
best fit but the expected slope for each fractal dimension
value $D_f$. As mentioned above, the correlation integral has
a well-defined power-law dependence but over a limited range:
for too high $r$ values the slope becomes smaller as $C$
approaches the asymptotical value $C=1$, while for too small
$r$ values statistical fluctuations produce a steeper slope.
This effect is more notorious for low fractal dimensions because
the low filling factor values allow a bigger random component when
placing the spheres in each level of hierarchy. Figure~\ref{mass3D}
shows the sphere mass $M(r)$ as a function of its radius for
the same fractal clouds shown in Figure~\ref{clouds}. Again,
the lines show the expected (theoretical) slope value. In this
case a low fractal dimension value ($D_f=1.5$) shows more
fluctuations because the fractal has more empty regions
(see Figure~\ref{clouds}) and there is a higher probability
of underestimating the mass. Notwithstanding the errors
being larger than with the correlation integral, a good
estimation of the fractal dimension can be done with this
method as it will be shown next. For a given (theoretical)
fractal dimension $D_f$, we have calculated $C(r)$ and
$M(r)$ for ten different random clouds and we have estimated
the average correlation ($D_C$) and mass ($D_M$) dimensions.
The results are plotted in Figure~\ref{both3D},
where we can see that both $D_C$ and $D_M$ are very similar to
the theoretical value $D_f$ in the analyzed range ($1 < D_f
< 3$), although the error bars are higher for $D_M$ mainly
for very low $D_f$ values where there are more empty spaces
in the simulated clouds.

\subsection{Projected Fractal Dimension}

\begin{figure}
\centering
\plotone{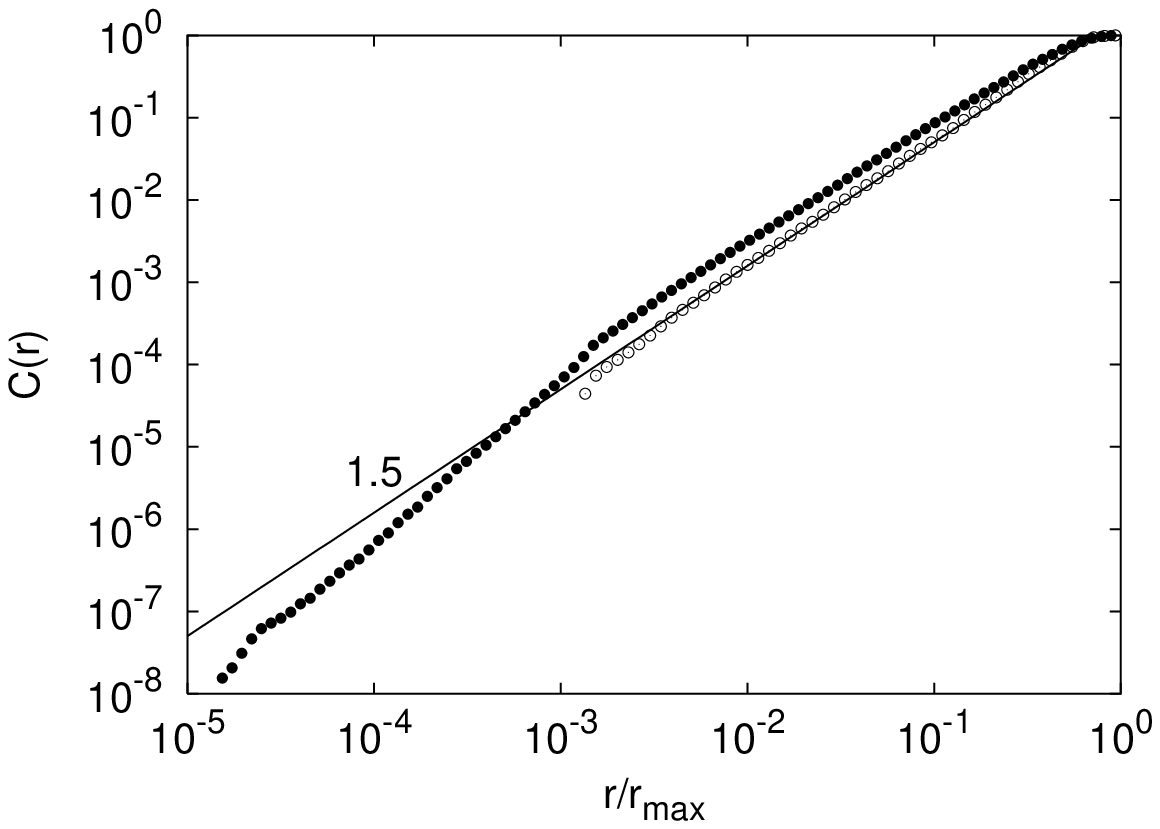}
\caption{The correlation integral $C(r)$ for the fractal
         with $D_f=1.5$ shown in Figure~\ref{clouds} (open
         circles) and for its projection (solid circles).
         The line shows the expected slope according to
         equation~(\ref{proyeccion}).}
\label{correlation2D}
\end{figure}

\begin{figure}
\centering
\plotone{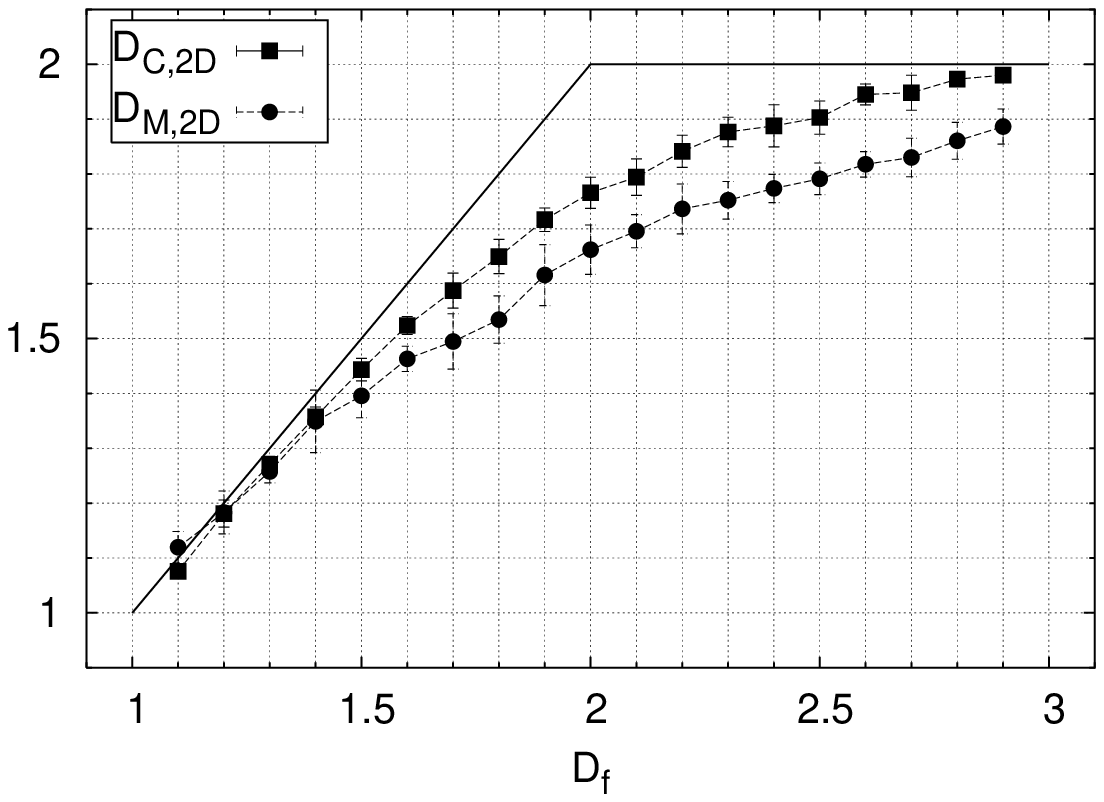}
\caption{The average correlation dimension ($D_{C,2D}$, squares)
         and mass dimension ($D_{M,2D}$, circles) for the projected
         fractals as a function of the tri-dimensional fractal
         dimension ($D_f$). The solid line shows the theoretical
         result given by equation~(\ref{proyeccion}).}
\label{both2D}
\end{figure}

We are interested in knowing what happens with the estimated
values of fractal dimensions when the measurements are made
on the two-dimensional projected images of the fractal clouds.
It has been proven analytically that the projection of a fractal
with tri-dimensional dimension $D_f$ yields a projected dimension
given by \citep{falc90,hunt97}:
\begin{equation}
\label{proyeccion}
D_{pro} = \min\{2,D_f\}\ . 
\end{equation}
Thus, $D_f$ and $D_{pro}$ are equal if $D_f < 2$ and
$D_{pro}=2$ if $D_f > 2$ (in this last case it would not
be possible to estimate $D_f$ from the projected image).
However, equation~(\ref{proyeccion}) was deduced for
indefinitely extensive (mathematical) fractals and its
validity for real finite fractals with certain random
component is not obvious. At least for random fractal
aggregates the equation~(\ref{proyeccion}) can not be
applied \citep{nels90,jull94,magg04}. We have made
projections of the simulated clouds on random planes
and we have calculated the average correlation and mass
dimensions for the resulting projections. An example is
shown in Figure~\ref{correlation2D},
where we have plotted the correlation integral for the
original tri-dimensional fractal and for its two-dimensional
projection using the cloud with dimension $D_f=1.5$ shown
in Figure~\ref{clouds}. The line indicates the expected
slope ($1.5$) for both the tri-dimensional and bi-dimensional
fractals (according to eq.~\ref{proyeccion}). For the assumed
values of $N$, $H$ and $D_f$ the smallest possible distance
between two particles in the simulated fractal is $r_{min}
\sim 10^{-3}$, and all the distances smaller than $r_{min}$
in the projected fractal arise from the random projection
itself. Thus, for $r > r_{min}$ we find that the projected
fractal has a slope in the log-log plot similar to (although
a little smaller than) the expected one, but for $r < r_{min}$
the slope increases tending to $\sim 2$ (the value corresponding
to a random distribution of particles on the plane). When
estimating the fractal dimension of the projected fractals
we have considered only distances greater than $r_{min}$.
Figure~\ref{both2D}
shows the measured correlation dimension ($D_{C,2D}$) and mass
dimension ($D_{M,2D}$) for the projected clouds as a function
of the tri-dimensional fractal dimension ($D_f$). Each point
is the result of calculating the average of ten different
realizations (random fractals) and the bars indicate the
standard deviations. We can see that the calculated values
are always below the theoretical result given by
equation~(\ref{proyeccion}), except for very low
($D_f \sim 1$) or very high ($D_f \sim 3$) fractal
dimension values. This result looks similar to the one
of \citet{magg04} obtained for the capacity dimension of
projected fractal aggregates. Since $D_{C,2D}$ varies
continuously with $D_f$, one might generally be able to
extract the fractal dimension of a three-dimensional cloud
by analyzing the correlation dimension of its projection,
even when $D_f > 2$, but $D_{C,2D}$ changes too slowly with
$D_f$ (within the error bars) for $D_f \ga 2.5$. In
spite of this degeneration both $D_{C,2D}$ and $D_{M,2D}$
could be used for doing a rough estimation of the fractal
dimension $D_f$.

\section{PERIMETER-AREA RELATION}
\label{perimeter}

The fractal dimension of a bounding contour can be determined
via the perimeter-area relation \citep{mand83}. In a plane,
the perimeter ($P$) and the area ($A$) are related by
\begin{equation}
\label{perimetro_area}
P \sim A^{D_{per}/2}\ ,
\end{equation}
where the dimension $D_{per}$ characterizes the degree of perimeter
contortion. Objects with smoothly varying contours (e.g. circles)
will have $D_{per}=1$ (i.e. $P \sim A^{1/2}$), whereas extremely
convoluted plane-filling contours will have $D_{per}=2$ ($P \sim A$).
Thus, the irregularity of a bounding contour is characterized
by $1 \leq D_{per} \leq 2$. The same argument applies for
three-dimensional surfaces but in this case the dimension is
$2$ for smooth surfaces tending toward $3$ for a convoluted
surface which fills the space. The contour of a projected
object is a subset of its surface and the measure of the
contour $D_{per}$ is not related (at least in an obvious way)
to, for example, the capacity $D_B$. The perimeter segmentation
reflects the roughness of the boundary while the capacity
measures the space-filling ability. Both of them, however,
give information on the fractal structure of the object.

We wish to investigate to which extent the information of the
tri-dimensional structure can be found from the projected
perimeter-based fractal dimension $D_{per}$. In order to do
this we first generate two-dimensional images applying the
following procedure: we project the simulated clouds on
random planes, then we place grids with various pixel sizes
(resolutions) whose ``brightness" is assigned by counting
the number of particles inside each pixel. We define the
image ``resolution" ($N_{pix}$) as the ratio between the
maximum two-pixel distance in the image and the pixel size
(i.e. the resolution is the maximum object size in pixel units).
The calculation of $D_{per}$ begins fixing a threshold
brightness level $n_{cri}$ and defining each ``object"
into the image as a set of connected pixels (having a
side or a corner in common) whose brightness value is
above $n_{cri}$. The area of each object is found by
counting the total number of pixels and the perimeter
is calculated by summing the lengths of the sides of
pixels along the edge of the object\footnote{In this
computation, we consider the sum of both the external
and the internal perimeter, i.e. the perimeter of the
internal voids are also computed}. Then, we calculate
the best linear fit in a $\log{P}-\log{A}$ plot using
the objects measured for different brightness levels and
the slope of this fit would be $D_{per}/2$. To increase
considerably the number of data points in the linear fit
we have taken $\sim 20$ brightness levels equally spaced
between the minimum and the maximum brightness. If an object
is relatively small (containing less than a few pixels)
its perimeter is not well defined and the structural
details will disappear, we therefore demanded that
objects contain a minimum ($\sim 15$) number of pixels.
Of course, several variations of the above procedure are
possible: the definition of connectivity for neighbouring
pixels (4-connected or 8-connected), the way for
calculating the perimeter (for example, the length
of the diagonal could be considered for the pixels
at corners instead of the sum of the two sides),
to take (or not) into account the inner holes,
or the minimum number of pixels that can constitute
one object in the fit. It has been proven that the
estimation of $D_{per}$ depends very weakly on these
criteria \citep{voge94}. In any case we tested the
influence of all these criteria and verified that the
result were not affected as long as objects too little
were not considered in the procedure.

\begin{figure}
\centering
\plotone{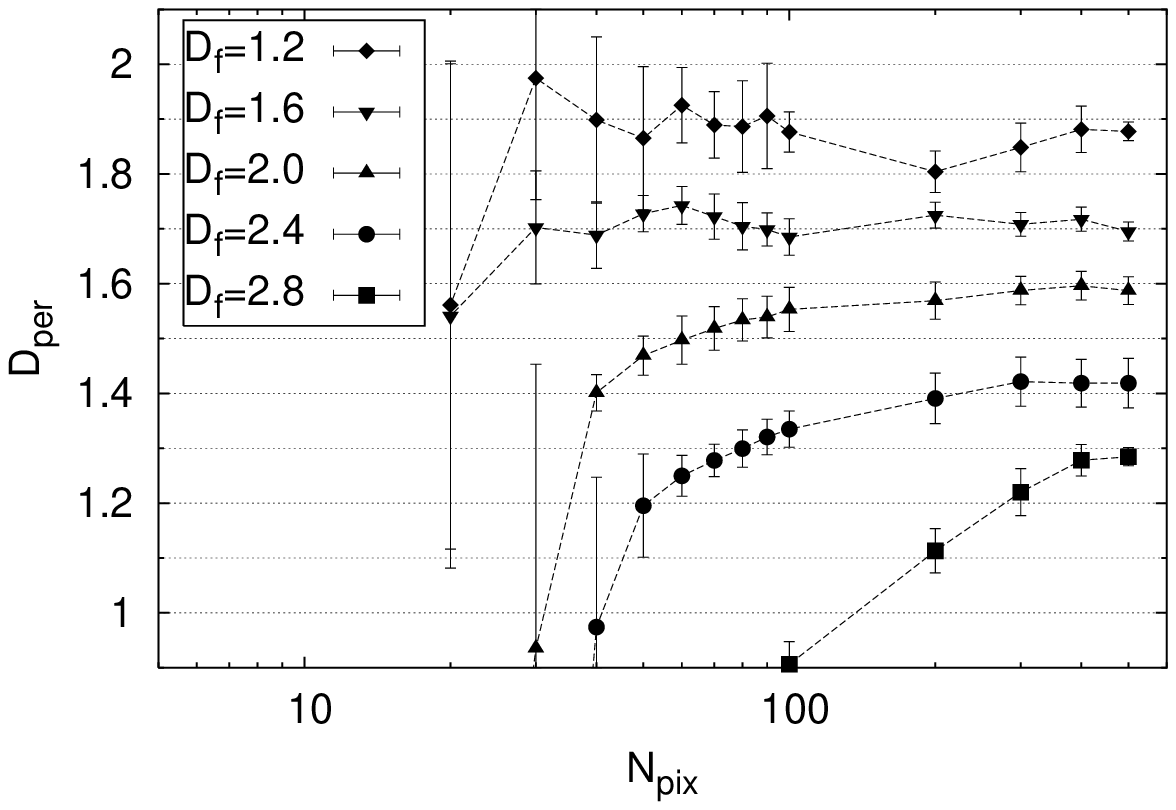}
\caption{The perimeter-area based dimension $D_{per}$ as a
         function of the projected image resolution $N_{pix}$
         for different cloud fractal dimensions: $D_f=1.2$
         (rhombuses), $D_f=1.6$ (inverted triangles),
         $D_f=2.0$ (triangles), $D_f=2.4$ (circles),
         $D_f=2.8$ (squares). The bars are the standard deviations
         resulting from ten different random realizations.}
\label{peri_reso}
\end{figure}

\begin{figure*}
\centering
\includegraphics[width=13.0cm]{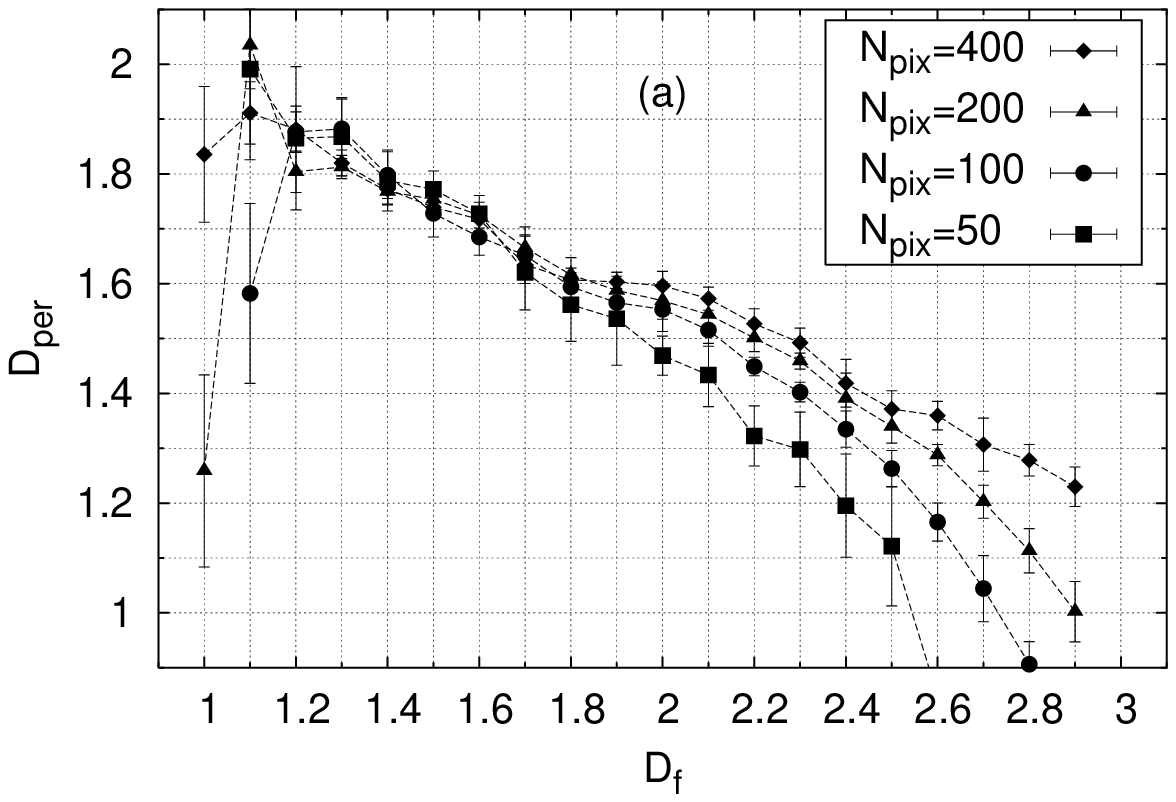}
\includegraphics[width=13.0cm]{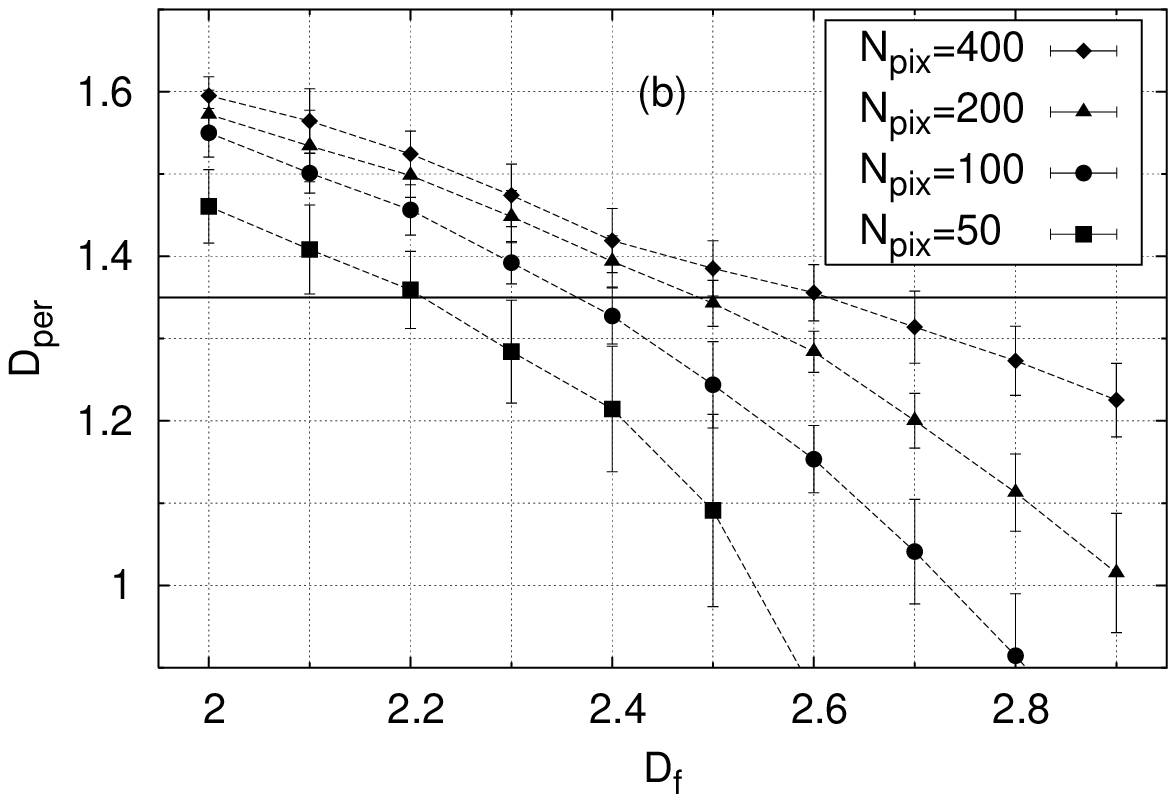}
\caption{The perimeter-area based dimension $D_{per}$ as a
         function of the cloud fractal dimension $D_f$ for
         different resolutions: $N_{pix}=400$ (rhombuses),
         $N_{pix}=200$ (triangles), $N_{pix}=100$ (circles)
         and $N_{pix}=50$ (squares); calculated for (a)
         10 different random clouds and (b) 50 random clouds
         but in a smaller range of $D_f$ values. The horizontal
         solid line shows the value $D_{per}=1.35$, and the bars
         are the standard deviations resulting from the different
         realizations.}
\label{peri_dime}
\end{figure*}

Figure~\ref{peri_reso}
shows the perimeter-area based dimension $D_{per}$ as
a function of resolution $N_{pix}$ for different values
of the cloud fractal dimension $D_f$ (for clarity, only
five values of $D_f$ are shown). The bars indicate the
standard deviations resulting from 10 different realizations
for each $D_f$ values, the mean errors of the best
linear fits in the $\log{P}-\log{A}$ plots always
were less than these error bars. The first feature we
note is that $D_{per}$ decreases as $D_f$ increases.
This is an expected result because bigger $D_f$ values
generate clouds with more round-shaped boundaries, otherwise
when $D_f$ decreases the clouds have more irregular
boundaries, as it can be seen in Figure~\ref{clouds}.
For a given fractal dimension $D_f$ there is a tendency
of $D_{per}$ to decrease as $N_{pix}$ decreases, and this is
because when the pixel size is bigger the details of the
roughness of the boundary disappear and the objects tend
to have smoother boundaries (and $D_{per} \rightarrow 1$).
Moreover, a very low number of pixels (and consequently
number of objects) and/or very low $D_f$ values (i.e. very
fragmented structures) increase errors notoriously.
On the other hand, at relatively high resolution
values $D_{per}$ converges toward some value which
we associate with the ``real" $D_{per}$ value resulting
from the projection. The dependence of $D_{per}$ on $D_f$
can be seen more clearly in Figure~\ref{peri_dime}
for different fixed resolutions (only a few $N_{pix}$ values
are shown for clarity, but for $N_{pix} \ga 400$ the curves
overlap each other within the error bars). In general $D_{per}$
decreases (from $\sim 2$ to $\sim 1$) as $D_f$ increases (from
$\sim 1$ to $\sim 3$), as can be seen in Figure~\ref{peri_dime}a,
but as resolution decreases the curves move down mainly at relatively
high $D_f$ values. Figure~\ref{peri_dime}b shows $D_{per}$ in a
smaller range of $D_f$ values but using 50 different random clouds
instead of 10, so that the random fluctuations become smoother (these
data are also shown in Table~\ref{tablaper}). We can estimate that if
the perimeter-area relation of interstellar clouds images yields
$D_{per} \simeq 1.35$ (this value is indicated with a horizontal line
in Figure~\ref{peri_dime}b) then the fractal dimension of the
tri-dimensional cloud should be some value in the range
$2.5 \la D_f \la 2.7$. This is higher than the result
$D_f \simeq 2.3$ that could be inferred when a relation
of the form $D_f = D_{per}+1$ is assumed.

\section{APPLICATION TO ORION A MOLECULAR CLOUD}
\label{orionA}

\begin{figure}
\centering
\plotone{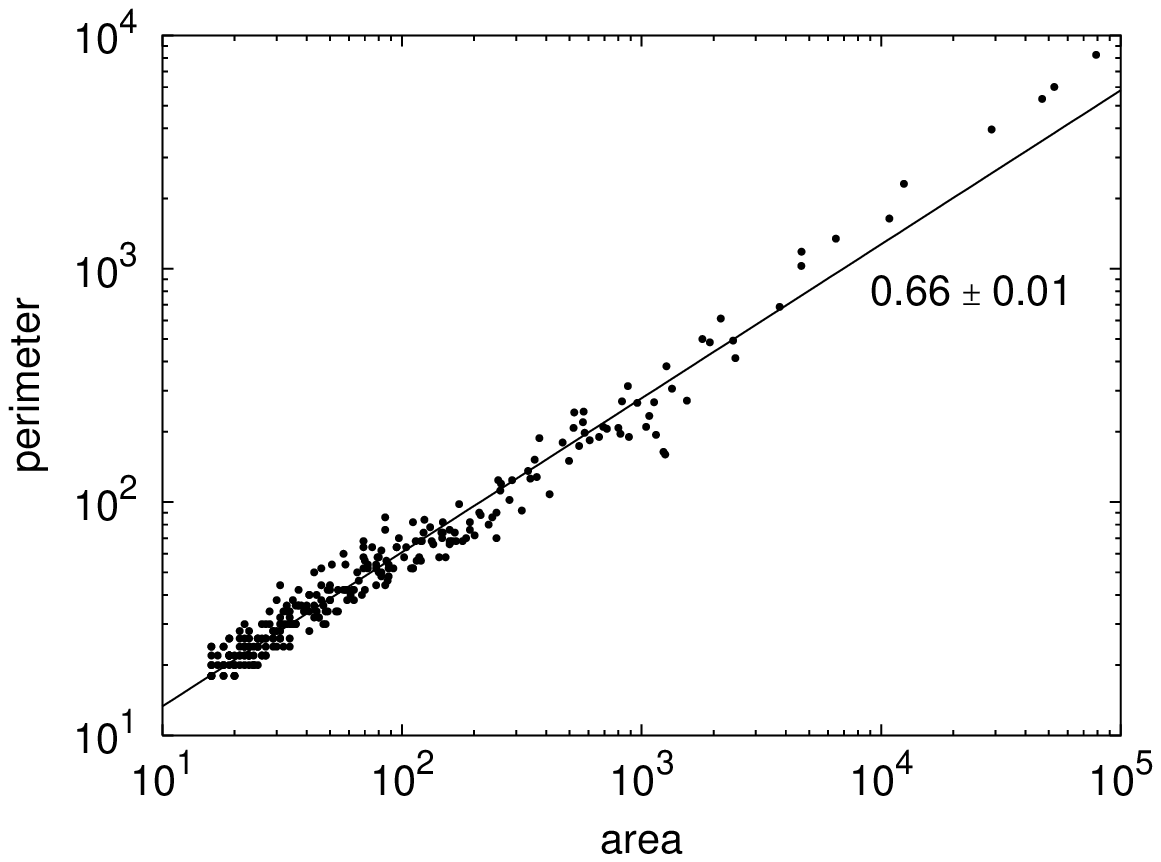}
\caption{The perimeter as a function of the area (in pixel
         units) for objects in the Orion A region at different
         intensity levels. The line is the best linear fit
         for all the points shown.}
\label{orion_peri}
\end{figure}

In this section we use the results obtained to estimate the
fractal dimension of Orion A giant molecular cloud\footnote{It
has to be mentioned that Orion A has a very elongated morphology,
and the analysis presented applies, in principle, to homogeneous
and isotropic fractals}. We use a $^{13}$CO~(1-0) integrated intensity
map obtained from high-resolution observations (15" beam and
40" grid spacing) with the 45~m telescope of the Nobeyama
Radio Observatory \citep{tate93}. To calculate $D_{per}$
we run exactly the same algorithm used in \S~\ref{perimeter}
for simulated clouds. Figure~\ref{orion_peri}
shows the perimeter-area log-log plot for Orion A (in pixel
size units). Each point represents different objects (clumps)
for different intensity thresholds and the line shows the best
linear fit including all the points plotted. The resulting
dimension (twice the slope) is $D_{per} = 1.32 \pm 0.02$,
a value within the range of measured dimensions in the
interstellar clouds. As concluded before for the simulated clouds,
from Figure~\ref{peri_dime}b (see also Table~\ref{tablaper}) one
obtains that the tri-dimensional fractal dimension of Orion A
should be around $2.6-2.7$.

\begin{figure*}
\centering
\includegraphics[width=17cm]{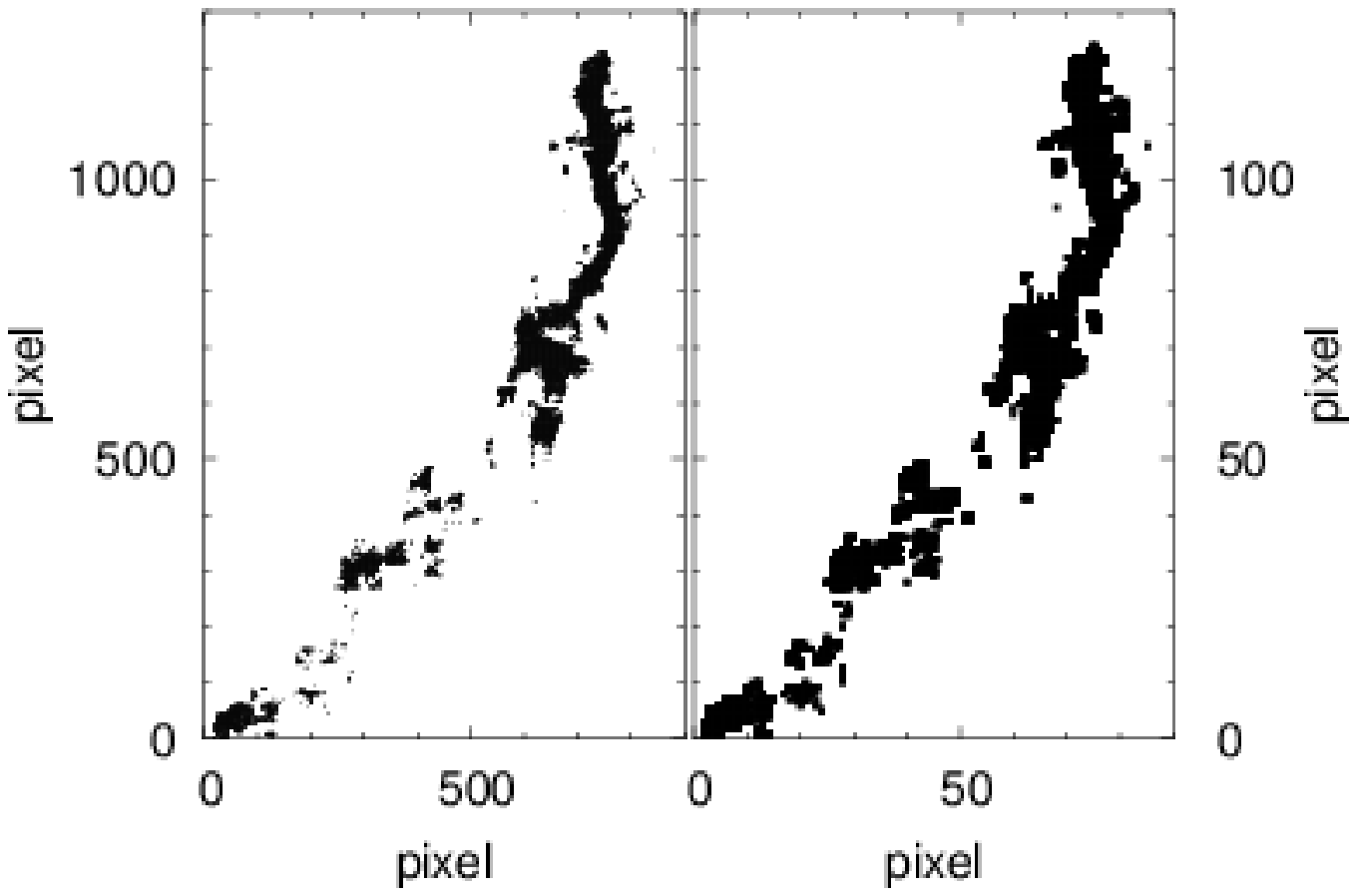}
\caption{Two images of the Orion A molecular cloud at
         different resolutions. Left panel: the original
         resolution given by \citet{tate93}. Right panel:
         ten times the original pixel length.}
\label{orion_imagen}
\end{figure*}

\begin{figure}
\centering
\plotone{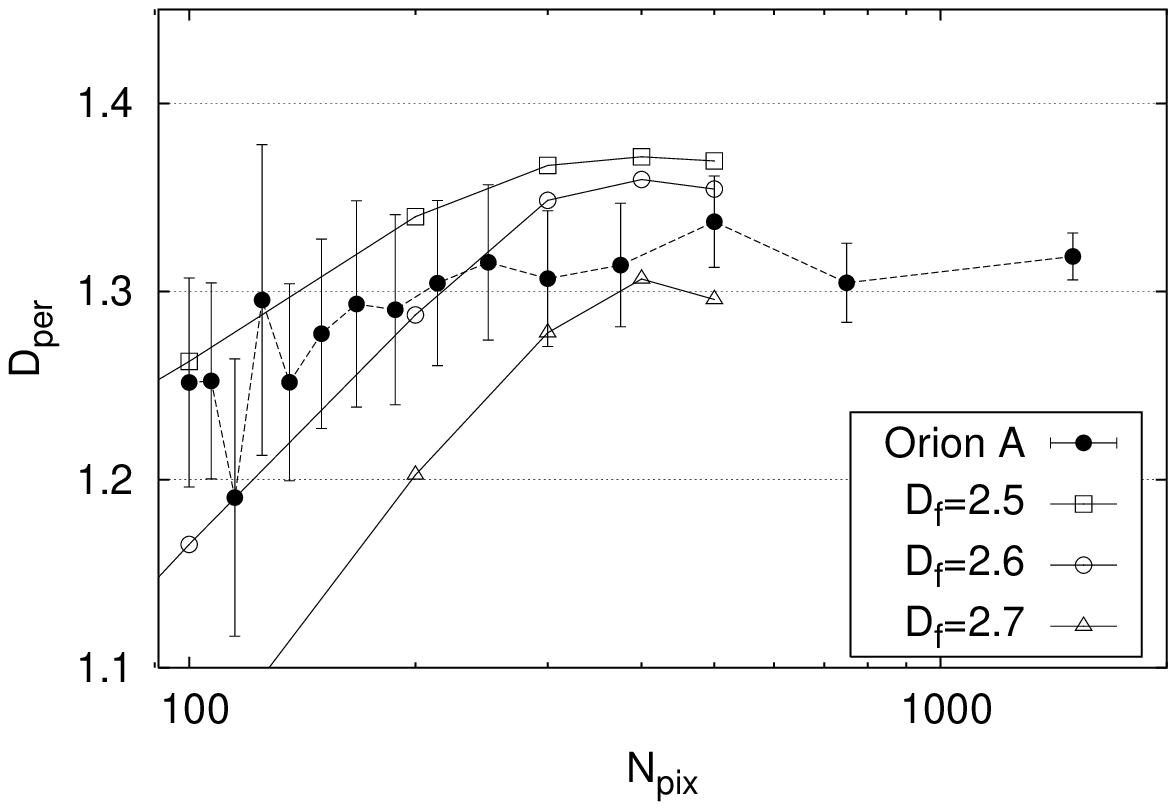}
\caption{The perimeter-based dimension $D_{per}$ as a function
         of the resolution $N_{pix}$ for Orion A (solid circles)
         and for simulated clouds with $D_f=2.5$ (open squares),
         $D_f=2.6$ (open circles) and $D_f=2.7$ (open triangles).}
\label{orion_reso}
\end{figure}

In order to verify this result we have decreased the resolution
of the Orion A image in successive steps, defining new larger pixels
whose intensities are given by the total intensities of the
neighbouring pixels. The left side panel in Figure~\ref{orion_imagen}
is an image with the original resolution given by \citet{tate93}
while in the right side panel the size of each pixel is ten times
the original one. The comparison of these images shows clearly
that progressively worse resolutions smooth the contours decreasing
its degree of convolutedness, because the details blend with each
other. Thus, $D_{per}$ should decrease when the resolution is
degraded beyond a critical value which depends on the fractal
dimension (as shown in Figure~\ref{peri_reso}). In each step we
have estimated the resolution $N_{pix}$ (number of pixels along
the maximum distance) and we have repeated the calculation of the
perimeter-based dimension $D_{per}$. The results are shown in
Figure~\ref{orion_reso},
where for comparison we have also plotted the calculated
results for the clouds simulated with $D_f=2.5$, $2.6$
and $2.7$. The bars indicate the standard errors of
the fits and, although they become relatively high at
low resolutions, it is clear from this Figure that the
way in which $D_{per}$ depends on $N_{pix}$ for Orion A
is consistent with a fractal dimension in the range
$2.5 \la D_f \la 2.7$. The dependence of $D_{per}$ on
$N_{pix}$ could be a better method to estimate $D_f$
that one single $D_{per}$ measure, mainly for low
resolution observations.

\begin{figure}
\centering
\plotone{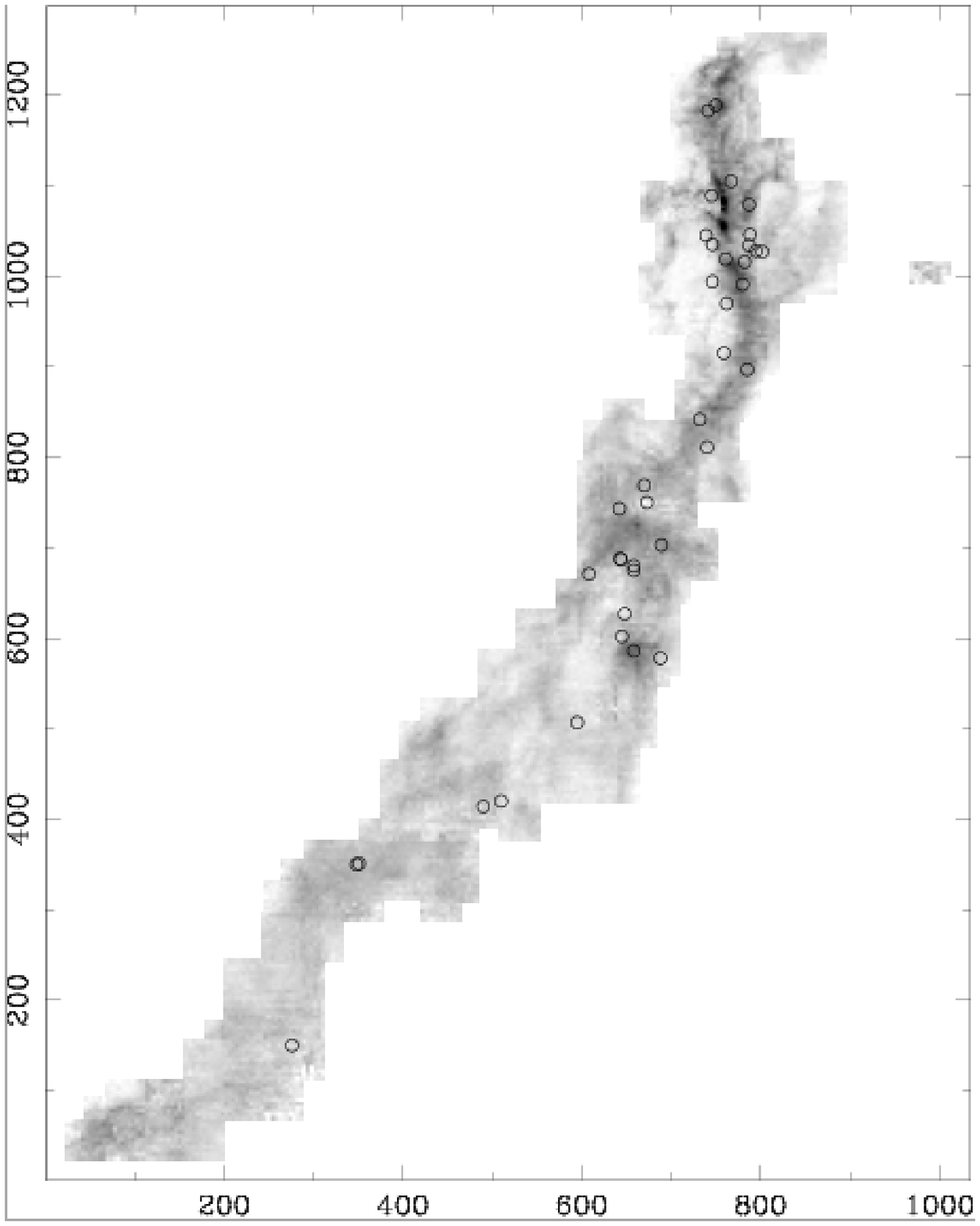}
\caption{
A $^{13}$CO map of Orion A molecular cloud. The open
circles indicate the positions of the sampling used to
estimate the mass dimension of this image.}
\label{enrique}
\end{figure}

\begin{figure}
\centering
\plotone{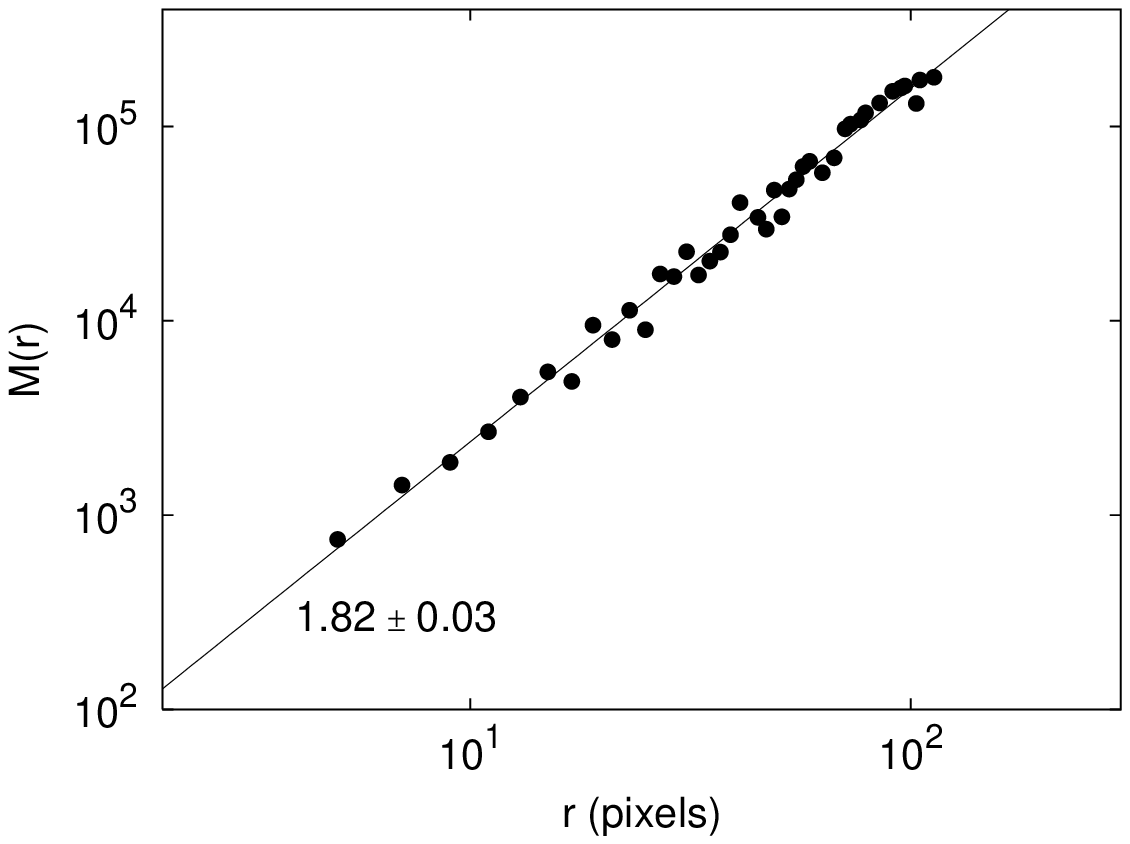}
\caption{The ``mass" $M(r)$ as a function of the cell size $r$
         for the Orion A molecular cloud. The solid line
         shows the best linear fit.}
\label{orion_mass}
\end{figure}

A final test we can do is to estimate the mass dimension
of Orion A by placing cells of different side sizes ($r$)
on the image and calculating the total intensity of the
cells. We call this quantity the ``mass" $M(r)$ assuming
that the intensity of each pixel is proportional to the
total column mass along the line-of-sight. We have used the
same algorithm explained in \S~\ref{theory} to calculate
the mass as a function of the cell size, doing random
sampling of relatively dense regions on the Orion A image.
We first generate $\sim 5000$ random positions along the
rectangle of $\sim 1000 \times 1300$ pixels containing
the Orion A image (see Figure~\ref{enrique}). The positions
located outside the image ($\sim 80\%$) are automatically
rejected as well as those positions where no relative
density maximum is found when increasing the size of
the cells (see \S~\ref{theory}). At the end of this
procedure, only $\sim 40$ points remained available
to be used in the calculation of the mass dimension.
Open circles in Figure~\ref{enrique}
indicate the position of the final sampling on the CO map of
Orion A used to calculate $M(r)$. We see that the algorithm avoids
to select low density regions where noise could introduce a higher
error in the determination of the mass dimension.
Figure~\ref{orion_mass}
is the resulting plot of $M(r)$ versus $r$ for the Orion A cloud.
The slope of the best fit ($1.82 \pm 0.03$) can be associated
with the mass dimension of the projected image ($D_{M,2D}$)
from which we can obtain, using the results shown in
Figure~\ref{both2D}, that the tri-dimensional cloud would
have a fractal dimension of $D_f \simeq 2.5-2.7$. This result
is in agreement with the one obtained above from the perimeter-area
relation.  We tested the reliability of this result varying the
sampling (both changing and increasing the number of random positions
on the image), and the estimated mass dimension remained unchanged
within the error bar resulting from the linear fit.

\section{DISCUSSION AND CONCLUSIONS}
\label{conclusions}

In this work we have generated tri-dimensional clouds
with given fractal dimensions ($D_f$) which we projected
on a plane to calculate the fractal dimension of the projected
image. {\it The main results derived from this empirical approach
concern the functional forms relating the tri-dimensional and
the projected fractal dimensions}. Both the correlation
and the mass dimensions increase as $D_f$ increases
(Figure~\ref{both2D}), but at least for the hierarchical,
finite, random fractals we generated, the projected
dimensions are always below the theoretical (expected)
values given by equation~(\ref{proyeccion}). We also
analyzed the perimeter-based fractal dimension ($D_{per}$)
and its dependence on both $D_f$ and the image resolution
($N_{pix}$). In general, $D_{per}$ is a decreasing function
of $D_f$, and it tends to decrease as $N_{pix}$ decreases
(Figures~\ref{peri_reso} and \ref{peri_dime}).

The results obtained with this method allow to estimate the
``real" fractal dimension of a cloud from its projection using
different fractal dimension estimators. The application to Orion
A yields $D_{per} \simeq 1.32$ for this molecular cloud which implies,
according to our results, a tri-dimensional fractal dimension
around $2.6-2.7$. Moreover, both the dependence of $D_{per}$ on
the resolution (Figure~\ref{orion_reso}) and the projected mass
dimension (Figure~\ref{orion_mass}) confirm that $D_f \sim 2.6$
for Orion A. This value is clearly higher than the result
$D_{per}+1 \simeq 2.3$ some times assumed in literature for
interstellar clouds \citep[e.g.][]{beec92}. \citet{elme96}
concluded that the size and mass spectra of interstellar clouds can be
the result of an ISM with {\it tri-dimensional} fractal dimension
around $2.3$, but \citet{stut98} have argued that this numerical
agreement could be coincidental. In fact, \citet{stut98} used
fractional Brownian motion structures to analyze the observed
ISM properties concluding that the index of the power spectrum
that best reproduces the observed characteristics is $\simeq 2.8$,
which would correspond to a cloud surface fractal dimension of
$D_{sur} \simeq 2.6$. Although this quantity is not necessarily
equal to $D_f$, it is interesting to note the agreement with our
results. The only result \citet{stut98} did not derive explicitly was
$D_{per} = D_{sur}-1 = 1.6$ for the corresponding perimeter-area
dimension.

The relationship between $D_f$ and the physical processes
determining the ISM structure is still an open issue.
Turbulent diffusion in an incompressible fluid gives
$D_f \sim 2.3$ \citep{mene90} and this fact has been
considered as evidence favoring the turbulent origin
for the observed fractal structure \citep{elme96}. Our
results seem to be in conflict with this interpretation,
but one should not forget that ISM is a complex system
where other effects (such as self-gravity, or
compressibility) have to be taken into account.
However, the results of the present study show
a clear evidence favoring fractal dimensions
for interstellar clouds higher than
the values more commonly assumed.

\acknowledgments

We want to thank Bruce Elmegreen for his valuable comments.
We also thank the referee for his/her helpful comments and
criticisms which improved this paper.
N. S. would like to acknowledge the funding provided by 
CONDES (Universidad del Zulia), FONACIT (Venezuela),
and the Secretaria de Estado de Universidades e
Investigacion (Spain).
E. J. A. acknowledges the financial support from
MECyD through grants AYA-2001-1696 and AYA-2001-3939-C03-01,
and from Consejeria de Educacion y Ciencia (Junta de
Andalucia) through the group TIC 101.
E.P. acknowledges financial support from
spanish grants AYA-2001-3939-C03-01 and AYA-2001-2089.

\clearpage
\begin{deluxetable}{ccccc}
\tablewidth{0pt}
\tablecaption{Calculated perimeter-area based dimension\label{tablaper}}
\tablehead{
\colhead{$D_f$} &
\colhead{$N_{pix}=50$} &
\colhead{$N_{pix}=100$} &
\colhead{$N_{pix}=200$} &
\colhead{$N_{pix}=400$}
}
\startdata
2.0 & 1.46$\pm$0.04 & 1.55$\pm$0.03 & 1.57$\pm$0.03 & 1.60$\pm$0.02 \\
2.1 & 1.41$\pm$0.05 & 1.50$\pm$0.02 & 1.53$\pm$0.04 & 1.56$\pm$0.04 \\
2.2 & 1.36$\pm$0.05 & 1.46$\pm$0.03 & 1.50$\pm$0.03 & 1.52$\pm$0.02 \\
2.3 & 1.28$\pm$0.06 & 1.39$\pm$0.03 & 1.45$\pm$0.03 & 1.47$\pm$0.04 \\
2.4 & 1.21$\pm$0.08 & 1.33$\pm$0.03 & 1.39$\pm$0.03 & 1.42$\pm$0.04 \\
2.5 & 1.09$\pm$0.12 & 1.24$\pm$0.05 & 1.34$\pm$0.03 & 1.39$\pm$0.03 \\
2.6 & 0.87$\pm$0.23 & 1.15$\pm$0.04 & 1.28$\pm$0.02 & 1.36$\pm$0.03 \\
2.7 & 0.65$\pm$0.28 & 1.04$\pm$0.06 & 1.20$\pm$0.03 & 1.31$\pm$0.04 \\
2.8 & 0.59$\pm$0.20 & 0.91$\pm$0.08 & 1.11$\pm$0.05 & 1.27$\pm$0.04 \\
2.9 & 0.50$\pm$0.39 & 0.78$\pm$0.09 & 1.02$\pm$0.07 & 1.23$\pm$0.04 \\
\enddata
\end{deluxetable}


\begin{thebibliography}{}
\bibitem[Bazell \& Desert(1988)]{baze88}
         Bazell, D. \& Desert, F. X. 1988, \apj, 333, 353
\bibitem[Beech(1987)]{beec87}
         Beech, M. 1987, \apss, 133, 193
\bibitem[Beech(1992)]{beec92}
         Beech, M. 1992, \apss, 192, 103
\bibitem[Buczkowski et al.(1998)]{bucz98}
         Buczkowski, S., Kyriacos, S., Nekka, F. \& Cartilier, L. 1998,
         Pattern Recognition, 31, 411
\bibitem[Chappell \& Scalo(2001)]{chap01}
         Chappell, D. \& Scalo, J. 2001, \apj, 551, 712
\bibitem[de Vega et al.(1996)]{deve96}
         de Vega, H. J., Sanchez, N. \& Combes, F. 1996, \nat, 383, 56
\bibitem[Dickman et al.(1990)]{dick90}
         Dickman, R. L., Horvath, M. A. \& Margulis, M. 1990, \apj, 365, 586
\bibitem[Elmegreen(1997a)]{elme97a}
         Elmegreen, B. G. 1997a, \apj, 477, 196
\bibitem[Elmegreen(1997b)]{elme97b}
         Elmegreen, B. G. 1997b, \apj, 486, 944
\bibitem[Elmegreen(2002)]{elme02}
         Elmegreen, B. G. 2002, \apj, 564, 773
\bibitem[Elmegreen \& Falgarone(1996)]{elme96}
         Elmegreen, B. G. \& Falgarone, E. 1996, \apj, 471, 816
\bibitem[Falconer(1990)]{falc90}
         Falconer, K. J. 1990,
         Fractal Geometry: Mathematical Foundations and Applications
         (London: Wiley \& Sons)
\bibitem[Falgarone \& Phillips(1990)]{falg90}
         Falgarone, E. \& Phillips, T. G. 1990, \apj, 359, 344
\bibitem[Falgarone et al.(1991)]{falg91}
         Falgarone, E., Phillips, T. G. \& Walker, C. K. 1991, \apj, 378, 186
\bibitem[Falgarone et al.(1992)]{falg92}
         Falgarone, E., Puget, J.-L. \& Perault, M. 1992, \aap, 257, 715
\bibitem[Ghazzali et al.(1999)]{ghaz99}
         Ghazzali, N., Joncas, G. \& Jean, S. 1999, \apj, 511, 242
\bibitem[Gonzato et al.(2000)]{gonz00}
         Gonzato, G., Mulargia, F. \& Ciccotti, M. 2000,
         Geophys. J. Int., 142, 108
\bibitem[Grassberger \& Procaccia(1983)]{gras83}
         Grassberger, P. \& Procaccia, I. 1983, \prl, 50, 346
\bibitem[Hetem \& Lepine(1993)]{hete93}
         Hetem, A. \& Lepine, J. R. D. 1993, \aap, 270, 451
\bibitem[Houlahan \& Scalo(1992)]{houl92}
         Houlahan, P. \& Scalo, J. 1992, \apj, 393, 172
\bibitem[Hunt \& Kaloshin(1997)]{hunt97}
         Hunt, B. R. \& Kaloshin, V. Y. 1997, Nonlinearity, 10, 1031
\bibitem[Jullien et al.(1994)]{jull94}
         Jullien, R., Tyhouy, R. \& Ehrburger-Dolle, F. 1994, \pre, 50, 3878
\bibitem[Khalil et al.(2004)]{khal04}
         Khalil, A., Joncas, G. \& Nekka, F. 2004, \apj, 601, 352
\bibitem[Kritsuk \& Norman(2004)]{krit04}
         Kritsuk, A. G. \& Norman, M. L. 2004, \apj, 601, L55
\bibitem[Larson(1992)]{lars92}
         Larson, R. B. 1992, \mnras, 256, 641
\bibitem[Maggi \& Winterwerp(2004)]{magg04}
         Maggi, F. \& Winterwerp, J. C. 2004, \pre, 69, 011405
\bibitem[Mandelbrot(1983)]{mand83}
         Mandelbrot, B. B. 1983, The Fractal Geometry of Nature
         (New York: Freeman)
\bibitem[Meneveau \& Sreenivasan(1990)]{mene90}
         Meneveau, C., \& Sreenivasan, K. R. 1990, \pra, 41, 2246
\bibitem[Nelson et al.(1990)]{nels90}
         Nelson, J. A., Crookes, R. J. \& Simons, S. 1990,
         J. Phys. D: Appl. Phys., 23, 465
\bibitem[Padoan et al.(2004)]{pado04}
         Padoan, P., Jimenez, R., Nordlund, A. \& Boldyrev, S. 2004,
         \prl, 92, 191102
\bibitem[Scalo(1990)]{scal90}
         Scalo, J. 1990,
         in Physical Processes in Fragmentation and Star Formation,
         ed. R. Capuzzo-Dolcetta, C. Chiosi \& A. Di Fazio
        (Dordrecht: Kluwer), 151
\bibitem[Soneira \& Peebles(1978)]{sone78}
         Soneira, R. M., \& Peebles, P. J. E. 1978, \aj, 83, 845
\bibitem[Stutzki et al.(1998)]{stut98}
         Stutzki, J., Bensch, F., Heithausen, A., Ossenkopf, V. \&
         Zielinsky, M. 1998, \aap, 336, 697
\bibitem[Tatematsu et al.(1993)]{tate93}
         Tatematsu, K. {\it et al.} 1993, \apj, 404, 643
\bibitem[Theiler(1987)]{thei87}
         Theiler, J. 1987, \pra, 36, 4456
\bibitem[Vogelaar \& Wakker(1994)]{voge94}
         Vogelaar, M. G. R. \& Wakker, B. P. 1994, \aap, 291, 557
\end{thebibliography}
\end{document}